%% file: paper.tex
\documentclass[sigconf,nonacm]{acmart}

\setcopyright{none}             % Suppresses standard commercial copyright
\makeatletter
\def\@copyrightspace{%
  \bgroup
  \insert\footins{%
    \bgroup
    \reset@font\footnotesize
    \interlinepenalty\interlookpenalty
    \splittopskip\vboxto\z@{}%
    \splitmaxdepth\maxdepth
    \vbox{%
      \hrule width \columnwidth height 0.4pt
      \vskip 5pt
      \noindent This work is licensed under a \href{https://creativecommons.org}{Creative Commons Attribution-NonCommercial-ShareAlike 4.0 International License}.
    }%
    \egroup
  }%
  \egroup
}
\makeatother

\newcommand{\code}[1]{\textbf{{\texttt #1}}}
\newcommand{\capt}[1]{\mdseries{\emph{#1}}}

\usepackage[T1]{fontenc}
\usepackage{xcolor}
\usepackage{soul}
\usepackage[english]{babel}

\usepackage[utf8]{inputenc}
\usepackage{subcaption}
\usepackage{graphicx}
\usepackage{tabularx}
\usepackage{adjustbox}
\usepackage{xspace} 
\usepackage{acronym}
\usepackage{mdwlist}
\usepackage{amsmath}
\usepackage{gensymb}
\usepackage{array}
\usepackage{enumerate}
\usepackage{paralist}
\usepackage{multirow}
\usepackage{amsmath}
\usepackage{url}
\usepackage{mdwlist}
\usepackage{balance}
\usepackage{enumitem}

\acrodef{NVM}[NVM]{Non-Volatile Memory}
\acrodef{PCM}[PCM]{Phase-Change Memory}
\acrodef{MCU}[MCU]{Microcontroller Unit}
\acrodef{LEO}[LEO]{Low Earth Orbit}
\acrodef{FRAM}[FRAM]{Ferroelectric RAM}
\acrodef{MRAM}[STT-MRAM]{Spin-Torque Transfer Magnetoresistive RAM}
\acrodef{ReRAM}[ReRAM]{Resistive RAM}
\acrodef{OBC}[OBC]{On-Board Computer}
\acrodef{PDM}[PDM]{Power Distribution Module}
\acrodef{VAE}[VAE]{Variational AutoEncoder}
\acrodef{DNN}[DNN]{Deep Neural Network}
\acrodef{SEU}[SEU]{Single Event Upset}
\acrodef{MBU}[MBU]{Multi Bit Effect}
\acrodef{SEFI}[SEFI]{Single Event Functional Interrupt}
\acrodef{SEL}[SEL]{Single Event Latch-up}
\acrodef{COTS}[COTS]{Commercial Off-The-Shelf}
\acrodef{ReLU}[ReLU]{Rectified Linear Unit}
\acrodef{NN}[NN]{Neural Network}
\acrodef{WBCE}[WBCE]{Weighted Binary Cross-Entropy}
\acrodef{KL}[KL]{Kullback-Leibler}
\acrodef{TP}[TP]{True Positive}
\acrodef{TN}[TN]{True Negative}
\acrodef{FP}[FP]{False Positive}
\acrodef{FN}[FN]{False Negative}
\acrodef{GAN}[GAN]{Generative Adversarial Network}

\newcommand{\fakepar}[1]{\vspace{0mm}\noindent\textbf{#1.}}
\newcommand\figref[1]{Fig.\,\ref{#1}}
\newcommand\secref[1]{Sec.\,\ref{#1}}

\theoremstyle{definition} 

\newcounter{glossy_enum}
\makecompactlist{list*}{list}

\setenumerate{topsep=0.2em,leftmargin=2.4em,itemindent=-0.0em}

\begin{document}

\title{Injection of Space Radiation Errors\\ on Low-power NVM using Generative Deep Learning}

\author{Eleonora Giudici}
\affiliation{\institution{Politecnico di Milano}\city{Milan}\country{Italy}}
\email{eleonora.guidici@mail.polimi.it}
\author{Antonio Miele}
\affiliation{\institution{Politecnico di Milano}\city{Milan}\country{Italy}}
\email{antonio.miele@polimi.it}
\author{Luca Mottola}
\affiliation{\institution{Politecnico di Milano}\city{Milan}\country{Italy}}
\affiliation{\institution{RISE and Uppsala University}\city{Uppsala}\country{Sweden}}
\email{luca.mottola@polimi.it}
\renewcommand{\shortauthors}{Giudici et al.}

\begin{abstract}
\input{abstract}
\end{abstract}

%\keywords{low-power NVM, space radiation, error-injection models, deep learning}
\maketitle

\input{intro}
\input{background}

\input{setup}
\input{injection}

\balance
\input{discussion}

\clearpage
\balance
\bibliographystyle{ACM-Reference-Format}
\bibliography{paper}

\end{document}

%% file: abstract.tex
We develop an error model for a \ac{NVM} device that represents the effects of physical faults experienced in a CubeSat mission in \ac{LEO}. 
The CubeSat carries an off-the-shelf low-power \ac{FRAM} chip and runs repeated read/write tests according to a predefined pattern. We compare every memory snapshot received from the satellite with a reference bitmap to identify bit errors. 
We visually analyze erroneous values distributions in the memory snapshots and we understand that they are frequently distributed according to regular spatial patterns. 
We define a generative deep learning technique to automatically learn these spatial patterns and to synthetically generate additional error patterns mimicking the observed ones. 
The model is capable at producing memory errors for general application-level experimentation. 
Our preliminary validation indicates that when comparing the model output with the data we collect, 0.92 of the generated bit flips correspond to bit flips in the memory snapshots and the model captures 0.91 of the observed bit flips. 
A comparison of the spatial distribution of bit flips in 10,000 generated error patterns also provides a Wasserstein distance of 0.073, indicating that the two distributions differ by roughly $7\%$ of the maximum possible displacement of their probability masses. 

%% file: intro.tex
\section{Introduction}
\label{sec:intro}

Nano-satellites are emerging as a new edge computing platform~\cite{orbitalEdge,berck2025launch,alen,paikowsky2017,denby2019orbital}. 
Unlike the monolithic satellite design that dominated space operations until the early 2000s, nano-satellites provide chea\-per operation and are robust to schedule variations and budget fluctuations, making it possible for academic institutions and even associations of hobbyists to design and deploy remote embedded sensing applications~\cite{narayana2020hummingbird,trusculescu2012nanosatellites,praks2011aalto,nagel2020nanosatellites,esper2000nasa,blackwell2012nanosatellites,zhang2026dynafilter}.

Nano-satellites are resource-constrained platforms, powered by solar energy harvesting and therefore subject to variable energy provisioning patterns~\cite{orbitalEdge,harvesting-survey}.
Their continuous operation is not guaranteed, as the limited energy budget is to be used for a number of different system functionality including computing, communication, and thermal control.
Systems are thus designed with \ac{NVM} to persist  application and systems states when the satellite is powered off and recharging energy buffers~\cite{denby2019orbital,paikowsky2017,berck2025launch}. 
Nano-satellites are also often exposed to high-energy, charged particles from sources like the sun, supernovae, and radiation trapped in planetary magnetic fields.
Such a complex mix of nuclear particles may affect the operation of the hardware onboard; in fact, their strike causes transient faults, such as bit flips, or even bear permanent effects, causing a circuit line or memory cell to be stuck at a fixed value.

CubeSats most often employ off-the-shelf hardware to reduce costs and, most importantly, to rely on the features and performance of modern hardware, as the space-rated one may lag several generations behind. 
Limited experimental data exists describing the effects of space radiation on off-the-shelf low-power components, for instance such as \acp{NVM}~\cite{dijks2026systematic,korkian2022single,oldham2013radiation}. 
Such a state of affairs hampers the development of error models to be used in application-level simulation that potentially enable large-scale application-level studies prior to actual deployments in orbit~\cite{classes}.
The issue two-fold: first, if we were to use the limited experimental data directly to perform error injection, the diversity of possible simulated executions would be extremely constrained; second, it would be exceedingly challenging to visually capture recurrent spatial and temporal data corruption patterns to design representative error models~\cite{classes,ruospo2022}.

\fakepar{A different take} This work is based on a CubeSat deployment in \ac{LEO} lasting about 3~months, which brings three \ac{NVM} chips  configured as scratchpad memory in orbit: a Fujitsu MB85RS64V \ac{FRAM}, an Everspin MR10Q010 \ac{MRAM}, and a Fujitsu MB85\-AS8MT \ac{ReRAM}. Throughout more than three months of  operation, the three \ac{NVM} chips are tested with a pre-determined read/wri\-te pattern. The collected memory snapshots are then eventually transmitted back to Earth and compared to a reference correct sequence to pinpoint bit errors due to space radiation. This analysis shows that bit errors may affect multiple consecutive bits, form error blocks of different sizes, and recur in different memory snapshots. An error model to be used for application-level error simulations in the early phases of a critical system design must therefore generate error patterns that preserve these aspects of the data.
  
To develop an accurate error model based on the experimental data collected, we take a different route compared with existing literature and let a deep learning pipeline capture recurrent spatial error patterns automatically.
We train a \ac{VAE}~\cite{kingma2014vae}, a type of generative model that uses a probabilistic approach to generate new data corruptions that mimic the training data.
We develop the model for the \ac{FRAM} chip, as that is  the setting where we have most data at disposal.
It would be possible, in principle, to apply the exact same techniques for the \ac{MRAM} and \ac{ReRAM} chip. 
We refrain from doing so because of the even more limited size of the data available for either memory, which would arguably prevent an accurate training.

Our preliminary validation indicates that when comparing the model’s output with the data we collect, 92\% of the generated bit flips correspond to bit flips in the memory snapshots, and the model accurately captures 91\% of the observed bit flips. An early comparison of the spatial distribution of bit flips in 10,000 generated error patterns reveals a Wasserstein distance of 0.073, indicating that the two distributions differ by approximately 7\% of the maximum possible displacement of their probability masses.

The rest of the paper unfolds as follows. \secref{sec:background} provides background information and positions this work with respect to related literate. \secref{sec:setup} describes the experimental platform and test procedures in orbit. Model development illustrated in \secref{sec:var}, whereas we report on a preliminary validation in \secref{sec:validation}. The paper ends in \secref{sec:end} with brief concluding remarks and future directions.

%The FRAM chip shows four distinct phases. In phases I, II, and III, the data contains error blocks of different sizes and consecutive error blocks are often separated by a multiple of eight bits. In phase IV, the vast majority of error blocks have size one. The model uses all four phases and can generate error patterns like those in the data, but it does not determine why the behavior of the FRAM chip changes.

%% file: background.tex
\section{Background and Related Work}
\label{sec:background}

Existing literature~\cite{pradhan1996} defines a \textit{fault} as a weakness or flaw in a system hardware component and the \textit{error} as the manifestation of a fault in the system state, for example, causing a deviation from the nominal behavior of program execution.
Faults and errors crucially refer to two different abstraction levels: the study of their relationship requires a critical cause-effect analysis across the two.

\fakepar{Radiation-induced faults} A \ac{SEU} is a transient fault in an electronic device caused by a single energetic particle striking a sensitive node, leading to a change in state in a single bit. 
A \ac{MBU} models the case when a single high-energy particle flips the data bits in two or more adjacent memory cells. 
%Unlike \ac{SEU}, which flips only one bit, an \ac{MBU} can cause more complex errors that are harder for conventional fault-tolerance techniques to handle.
When radiations hit control logic, instead, they may cause a \ac{SEFI}, which leads a device to enter a temporary non-functional state.
A \ac{SEFI} is ``non-destructive interruption resulting from a single ion strike that causes an electronic component to reset, hang, or enter a different operating condition or test mode''~\cite{jesd57}.
\acp{SEFI}  are usually set apart from \acp{SEU} and \acp{MBU}, as they persist until the system is reset, whereas the latter may be corrected by simply rewriting memory locations.
%may occur whenever in a program execution, and disappear also without powering off or resetting the system.

Due to their transient nature and what is generally known as the ``self-healing'' effect~\cite{oldham2003total,schwank2008radiation}, \acp{SEU}, \ac{MBU}, and \acp{SEFI} are generally referred to as \emph{soft errors}; on the other hand, radiations can cause also permanent effects, such as \acp{SEL}, which lock a memory cell to a fixed value. 
In contrast to soft errors, radiation-induced permanent faults are also called \emph{hard errors}.

\fakepar{Radiations effects in \ac{LEO}}
Galactic cosmic rays and solar cosmic rays are the two main sources of radiations in \ac{LEO}. 
Their interaction with the Earth magnetic field and the atmosphere, causes the so-called Van Allen radiation belts~\cite{Kovar2020}. 
Radiation intensity in \ac{LEO} varies with the specific position around the Earth and also due to specific events such as solar flares~\cite{noaa}.
First verified evidences of cosmic radiation causing \acp{SEU} in digital circuits come from observations in first generation satellites during the 1970s~\cite{binder1975}.
Further evidences appear in all subsequent missions up to the recent time, as in SpaceX satellites~\cite{spacex}. 

Using off-the-shelf hardware instead of space-rated hardware prompted efforts to understand the effects of space radiations in \ac{LEO}.
\citet{6353715} deploy a Virtex-4 FPGA in \ac{LEO}.
They use a digital signal processing application as a benchmark across two different FPGAs, a larger XQR4VLX200 and smaller XQR4VSX55.
They focus on where in orbit \acp{SEU} and \acp{MBU} occur, and at what rates.
They find that the rates are proportional with FPGA size.
\citet{yaacoub2025fault} deploy an Arty Z7-20 board in \ac{LEO}, which combines programmable logic resources similar to traditional FPGAs with a dual-core ARM Cortex-A9 SoC.
They compare the execution of a Fast Fourier Transform implementation on the FPGA chip and on the ARM core to test the validity of existing fault tolerance mechanisms, including triple-modular redundancy.
The single fault model is found to be valid in their setting.
\citet{1035377} compare the use of an RH-3000 radiation-hardened processor against an IDT-3081 off-the-shelf processor in \ac{LEO}.
The off-the-shelf processor is an order of magnitude faster, yet it experiences two orders of magnitude more errors~\cite{1035377}.
Common to these works is the goal of understanding the effects of space radiation on off-the-shelf hardware, but we focus \ac{NVM} instead.

\fakepar{Fault and error models} 
In the study of fault detection and handling techniques, developers frequently need to perform fault injection experiments~\cite{pradhan1996}. 
They adopt a \textit{fault model} to that end, that is, a formal description of physical failures that may occur in hardware, and use it in emulation or simulation to systematically and deliberately corrupt the system, or a representation of it.
By doing so, they analyze how the errors propagate through an execution and thus whether a given technique works effectively.

Fault injection may be too onerous or time-consuming to be adopted; for example, due to the complexity of the hardware platform, the unavailability of a corresponding simulation model, or because legacy components are employed and therefore internally inaccessible. 
In these cases, developers may opt to corrupt a running program by directly injecting errors, for example, by corrupting program variables.
To do so, they need an \emph{error model}.

While consolidated literature on fault models exists, error modeling is more challenging~\cite{classes,ruospo2022}. 
Errors indeed may manifest in a different location than the related fault, and in a highly more articulated way.
The different levels of abstractions corresponding to faults and errors, as described earlier, indeed offer different standpoints.
Many times, a program may also be observed only in a fixed set of locations; as a consequence, the original fault may propagate and expand its effects through a program's execution in ways that cannot be traced accurately.
An error model may thus depend on both the hardware, as the related fault does, and on the  functionality executed by the program where the fault occurs.
Due to the limitations of the CubeSat deployment, our investigation here centers on \emph{errors} as they manifest on the \ac{NVM} chips.

Existing literature on error modeling close to our work include, for example, the work of \citet{ruospo2022}, who investigate of the effects of radiations in the DRAM storing the weights of \acp{DNN}. They analyze how corrupted values are spatially distributed in the memory grid, and define error models capable at capturing these spatial patterns and their temporal rates. 
They thus design an error simulator capable of corrupting the weights of the executed \ac{DNN}, with the aim of evaluating the robustness of these applications.
This contribution is close to out work since both analyze the errors causes in memory snapshots aiming at extracting regularities in the distribution of corrupted values.

\citet{classes} present an investigation of spatial corruption patterns in the context of GPUs running \acp{DNN}.
They define error models at the granularity of single \ac{DNN} layers by injecting faults in GPU logic during the execution of single layers, and analyzing the corruptions in the output tensor. 
The outcome is a large set of error models, one for each class of \ac{DNN} layer, characterized by specific spatial patterns and occurrence rates. 

The main limitation of these works is that spatial and temporal error patterns are derived visually by developers by manually inspecting a large number of corrupted memory snapshots.
We explore a different approach and let a \ac{VAE} capture these patterns automatically during training; then, we use the \ac{VAE} to generate new data corruptions mimicking the training data. 
We thus eliminate the necessity of a time-consuming manual activity and increase the quality of the output, as the \ac{VAE} can capture more general corruption patterns that may not be seen by visual inspection.

%% file: setup.tex
\section{CubeSat Mission}
\label{sec:setup}

This section presents the CubeSat mission performed for collecting experimental data about how radiations affects \ac{NVM} devices in space.
We describe the characteristics and features of the space vehicle we deploy, the memory array under test and the test procedure, and provide key information about how the mission unfolded.

\subsection{Space Vehicle}

\begin{figure}[t]
    \centering
    \hspace{-10pt}
    \subfloat[CubeSat platform.\label{fig:cubesat_prototype}]{
        \includegraphics[width=0.15\textwidth]{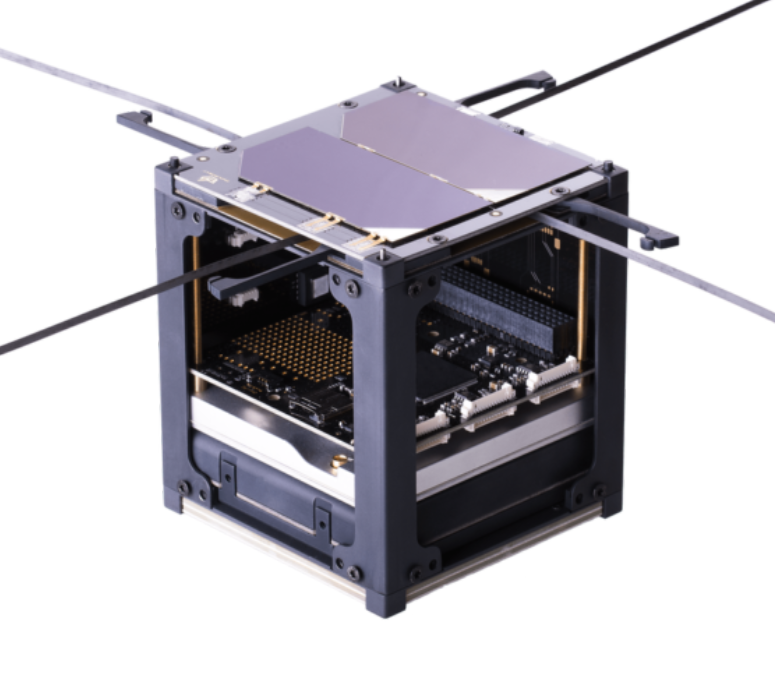}
    }\quad
    \subfloat[Layered architecture.\label{fig:cubesat_prototype2}]{
        \includegraphics[width=0.22\textwidth]{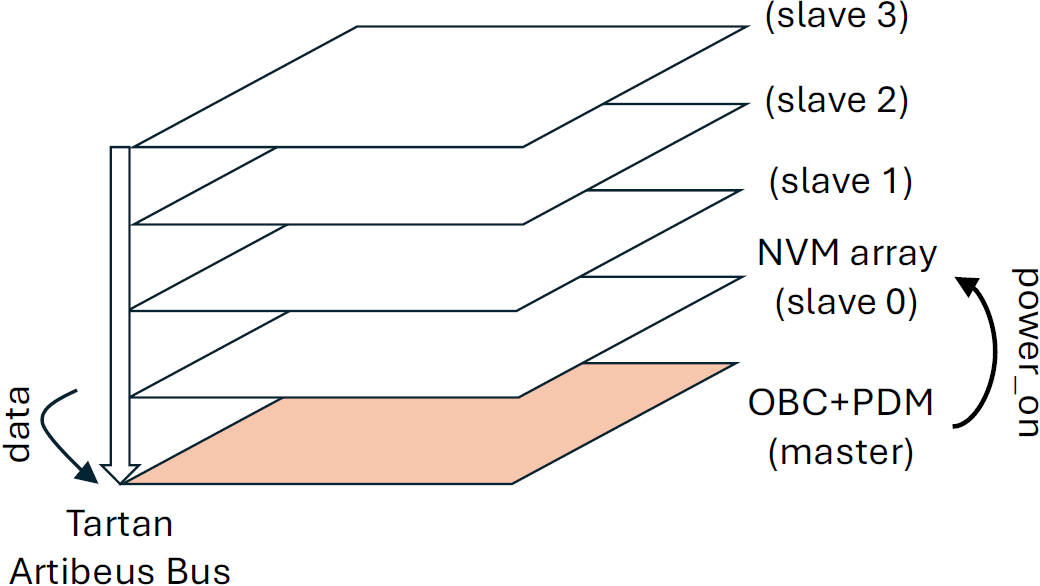}
    }
    \vspace{-1mm}
    \caption{Space vehicle and hardware architecture. \capt{The CubeSat hosts the  \acs{OBC} and \acs{PDM} at the bottom, tasked with mission control and energy management. The slave $0$ hosts the \ac{MCU} controlling the experiments and the three \ac{NVM} chips we test.}}
    \vspace{-3mm}
    \label{fig:cubesat_overview}
\end{figure}

We deploy a CubeSat space device, shown in \figref{fig:cubesat_prototype}, built on the 1-Unit CubeSat platform developed by EnduroSat \cite{Endurosat}.
The 1-Unit form factor is the smallest for this class of space devices.  
Because of the limited physical size, the internal layout and placement of components is extremely optimized, and only roughly 14\% of the internal space remains unused.

\begin{table}[tb]
      \vspace{-3mm}
  \centering
    \footnotesize	
    %\small
    \begin{tabular}{|c|c|}
    \hline
    \textbf{NVM chip} & \textbf{Fujitsu FRAM MB85RS64V}\\
    \hline
    Bit configuration & 8,192 words × 8 bits  \\
    \hline
    Endurance & $10^{12}$cycles/byte \\
    \hline
    Retention & $>10$ years \\
    \hline
    Supply voltage (min–max) & 3.3–5.5 V \\
    \hline
    Current draw (min–max) & 10µA-1.5mA \\
    \hline
    \end{tabular}
    \caption{\ac{NVM} chip specifications. \capt{The information is taken from the corresponding datasheets.}}
    \vspace{-5mm}
    \label{tab:datasheet}
\end{table}

The onboard hardware follows a layered master-slave architecture, as illustrated in \figref{fig:cubesat_prototype2}. The device located at the base functions as the satellite’s master \ac{OBC}. 
It is built around a space-rated version of IBM’s 6x86 CPU, a 32-bit computing core featuring a superpipelined architecture and a hardware floating-point unit.
Despite being based on an almost 30-year-old design, its space-rated version is still used in space operations, as the software written for it underwent extensive formal verification and proved reliable across multiple missions~\cite{stakem2004migration}.
A dedicated radiation-strength 10 mm aluminum shield separates each of the slave devices from each other and from the \ac{OBC}.

Besides logging of mission data and general bookkeeping, the software aboard the \ac{OBC} controls a custom \ac{PDM} also at the base. 
The \ac{PDM} uses the energy coming from four CTJ30 CESI solar cells \cite{cesi} to power the \ac{OBC}. 
%The panels feature up to 29.5\% efficiency and offer the largest possible effective cell area for 1-Unit CubeSats, ultimately providing up to 2.4 W per panel in LEO. 
Any leftover energy is stored in a soft-reconfigurable supercapacitor array on the \ac{PDM}, which can vary total capacitance according to the net input power. 
This feature is crucial in the operation of a resource-constrained nano-satellite, as input power from the solar cells may vary widely as the orbit unfolds over time.

The \ac{OBC} instructs the \ac{PDM} to supply energy to one or more of the four slave devices onboard. 
Independent parties program these devices to perform various experiments.
The \ac{OBC} determines what device(s) to power depending on their individual energy figures and the amount of experimental data output up to a given point, in an attempt to ensure fairness of energy allocations. 
The slave devices relay data to the \ac{OBC} through a simplified version of the Tartan Artibeus Bus~\cite{denby2022tartan} using a serial line.

The UHF Transceiver II module from EnduroSat enables downlink communication with Earth by leveraging the SatNOGS~\cite{Satnogs} global network of satellite ground stations for data retrieval. 
Communication occurs using space-proven FCC protocols and extreme data redundancy. Corrupted packets are filtered out by the SatNOGS network as soon as they fail one of the multiple redundancy checks employed. 
This guarantees that the errors we observe occur on the satellite and not during transmission back to Earth.

Our design the experimental isolates the origin of observed errors from other satellite-level effects. 
The \ac{OBC} reports internal-temperature telemetry throughout the mission, and we found no recorded value outside the operating range of any of the three \ac{NVM} chips. In addition, the \ac{OBC} power-conditioning circuitry filters supply surges before power reaches the slave devices. These controls specifically address thermal and power issues, which are known to affect the behavior of computing systems in orbit~\cite{li2024exploring,xing2024deciphering}  . Together with the physical isolation of the experimental scaffold and the pre-launch radiation tests, they support attributing the observed errors primarily to radiation effects on the \ac{NVM} chips. 
This attribution does not, however, determine whether an individual radiation-induced error originates in a memory cell or in on-chip peripheral circuitry, which can itself be radiation-sensitive~\cite{bosser2018single}.

Due to limitations of the Tartan Artibeus Bus \cite{denby2022tartan} and of the way the \ac{PDM} distributes available energy, it is not possible to accurately correlate the collected data with absolute positioning or global time, even though this information is available at the \ac{OBC}.
A different design where this is possible, on the other hand, would not fit the physical size of a 1-Unit CubeSat.
As a result, we cannot make use of temporal or positional data.

\subsection{Memory Array and Operation}
\label{sec:experiments}

The experimental platform hosts three \ac{NVM} chips, that are: 
\begin{inparaenum}[i)]
\item Fujitsu \ac{FRAM} MB85RS64V\footnote{\url{https://www.fujitsu.com/uk/Images/MB85RS64V.pdf
}},
\item Everspin \ac{MRAM} MR10Q010\footnote{https://www.everspin.com/sites/default/files/MR10Q010\_Datasheet.pdf}, and
\item Fujitsu \ac{ReRAM} MB85AS8MT.\footnote{https://www.ramxeed.com/assets/images/products/datasheet/ReRAM/MB85AS8MT-DS2v2-E.pdf}
\end{inparaenum}
We chose these technologies to test the robustness of different options that, because of their energy figures and size, could be deployed on a resource-constrained nano- or pico-satellite.
We do not consider other \ac{NVM} technologies, such as Flash and \ac{PCM}, because of high energy consumption or the early stage of qualification for space missions. 
The modeling effort described next only considers the \ac{FRAM} chip, as insufficient data was collected from the other two \ac{NVM} chips.

\begin{figure}[t]
    \centering
        \vspace{-1mm}
    \includegraphics[width=0.18\textwidth]{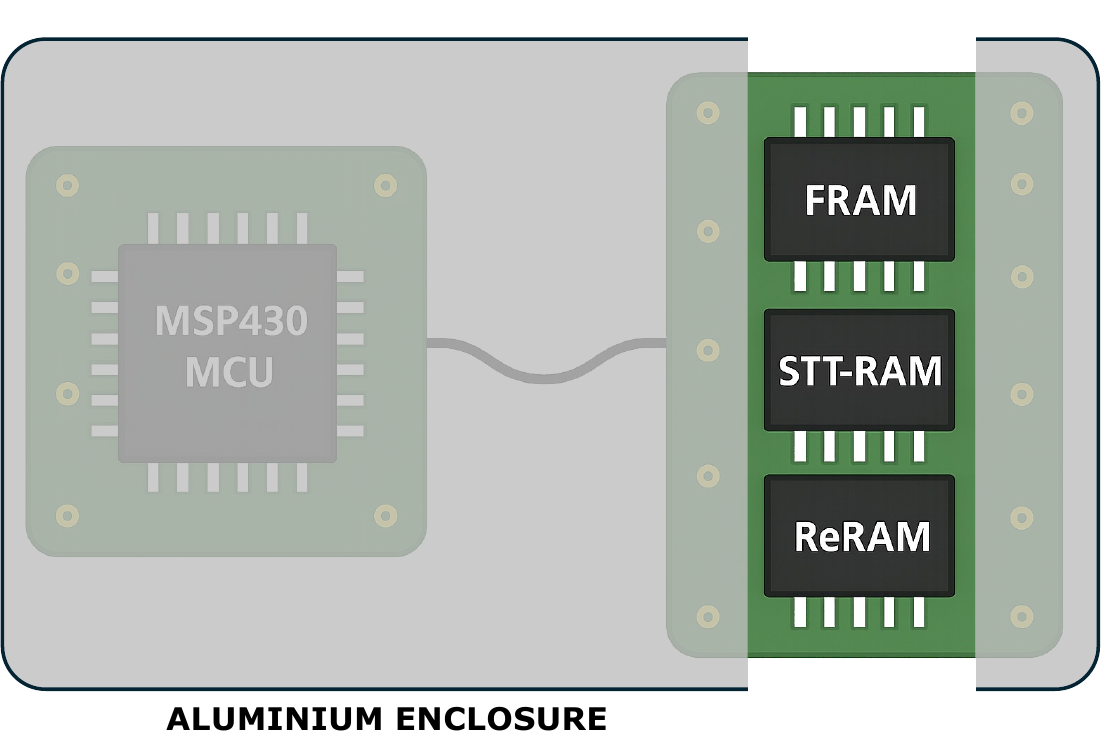}
        \vspace{-1mm}
\caption{Slave 0 device on the CubeSat with \ac{NVM} array. \capt{An additional 14mm-think aluminum enclosure protects the experimental scaffold, in addition to the 10 mm aluminum shields between slave devices/OBC, and the outer aluminum alloy frame of the satellite.}}
    \vspace{-3mm}
    \label{fig:satPCB}
\end{figure}

\fakepar{Hardware layout} 
The slave $0$ device in \figref{fig:cubesat_prototype2} includes two boards, whose relative layout is shown in \figref{fig:satPCB}.
One board hosts a TI MSP430FR6989 \ac{MCU}, which acts as the local controller for the experiment. 
The second board hosts the three NVM chips under test and is connected to the MCU through SPI. 
The physical separation mirrors the logical separation in the experiment: the MCU board implements the experimental scaffoldwhereas the memory board hosts the \ac{NVM} chips we test. 

%The boards are kept in a 14 mm aluminium enclosure that provides additional isolation from radiation on top of what is already provided by the outer aluminium alloy frame of the CubeSat and by 10 mm aluminium shields between the slave devices and the OBC, mentioned earlier. 
%The additional shielding provides a safeguard for our experiments, ensuring that the errors we observe are the result of faults occurring on the \ac{NVM} chips and not on the experimental scaffold around them. 

Both boards are placed inside a 14 mm aluminum enclosure that provides additional shielding \emph{beyond} the 10 mm aluminum shields sitting between each of the slave devices and the OBC, and the outer aluminum alloy frame of the satellite. 
This enclosure is designed to protect the experimental scaffold. 
This choice is central to our experiments: the additional protection reduces the probability that the MCU, board-level interconnects, or other elements of the experimental scaffold corrupt the memory snapshots before they are relayed to the OBC.
We experimentally verify this as explained next, prior to the actual launch.
Crucially instead, in terms of exposure to space radiation, the \ac{NVM} chips are placed \emph{in the same conditions} as they would be on any other application-specific nano- or pico-satellite platform~\cite{denby2022tartan,pocketcube, POGHOSYAN201759}. 

In essence, this strategy is an attribution mechanism rather than a deployment recommendation.
This kind of shielding is indeed affordable (with some effort) on a 1-Unit CubeSat operating as an orbital experimental testbed but would be likely infeasible in a real application.
Allocating 14 mm of aluminum around a small experimental board would be difficult to justify, because the same volume and mass would otherwise be used for mission payloads, sensors, or application-specific subsystems~\cite{praks2011aalto,blackwell2012nanosatellites}. 

%We test this design prior to launching the CubeSat with an Americium/Beryllium (AmBe) source with an activity of 2.6 GBq that generates a yield of 156,000 neutrons per second, and an Amptek ``Mini-X'' miniature x-ray source with silver target at 30kV and 200uA and no collimator, which yields a dose rate of $32.4 \mu Sv/h$ with a flux of $2.6 * 10^9 counts/second/mm^2$, and find no evidence of any internal radiation leakage.\am{se aggiungi la sezione di confronto dei risultati ottenuti nei radiation test con quelli in orbita qui metterei una frase che dice che abbiamo utilizzato questi dati per un'analisi comparativa.}

%%%%%LUCA: note for myself
\fakepar{Experiment design and execution} To build an error model, we seek to characterize the \ac{NVM} chips themselves in the presence of space radiation, not to assess whether a specific application produces correct results. 
At the same time, we must reduce the risks that, despite the additional shielding, radiation may impact other stages of the processing besides the \ac{NVM} operation.

We call \emph{active cycle} the uninterrupted period of time the \ac{OBC} decides to power the slave $0$ device, which produces a variable number of memory snapshots depending on the cycle duration.
This is the same notion of active cycle in regular intermittent computing~\cite{ahmed2024internet}.
Continuously during an active cycle, for each \ac{NVM} chip, the \ac{MCU} first zeroes the memory; next, it writes a specific bit string progressively obtained from a simple sequence of bit-wise operations that commence from a \emph{single 16-bit seed}, stored in the built-in \ac{FRAM} of the \ac{MCU}.
%The string is as large as the target \ac{NVM} chip and includes the same number of zeros and ones. 
Writes occur through the \code{SPI\_WRITE\_CMD} command in blocks of 256 bits.
Next, the \ac{MCU} reads back the memory content again using SPI and progressively hands it over to the \ac{OBC}, where it is queued for transmission. 
%As these steps are not instantaneous, what we eventually observe are indeed errors on the \acp{NVM}, as we have no information as to when the related fault(s) occurred.
Read operations occur immediately after write operations while the \ac{MCU} remains powered; thus, retention times are not expected to impact the results.

Note that this design is more robust than, for example, storing the entire bit sequence on the built-in \ac{FRAM} and merely copying it over to the \ac{NVM} chips.
The built-in \ac{FRAM} is limited in size, and some additional processing would be required anyway to operate on the larger \ac{MRAM} and \ac{ReRAM}.
Errors on the original bit stream on the built-in FRAM may, indeed, become indistinguishable from errors on the external \ac{NVM} chips.
Instead, as the operations the MCU performs to generate the bit string to be written on the \ac{NVM} chips are known, we can detect errors in the 16-bit seed by looking for a specific error pattern appearing \emph{simultaneously on all} the external \ac{NVM} chips. 
We calculate that the probability of this error pattern appearing for reasons other than an error in the seed is $<10^{-18}$. 
In the data we collect, this error pattern never occurs. 

This procedure gives us direct bit-level visibility into every memory location eventually returned from the satellite. 
Running actual application logic would make attributing data errors much harder; for example, later processing stages could overwrite or absorb a radiation-induced fault, or skew the analysis toward an application-specific workload. 
Our bit-level measurements identify the error patterns that the model should inject, whereas the semantics of the affected data ultimately determine whether an error is masked, causes bounded degradation, or becomes a persistent failure.
In this sense, the procedure we adopt provides a foundation for constructing the error model, making it available for general application-level error simulation experimentation. 

\subsection{Mission}
\label{sec:mission}

Before launch, we validated the hardware and experiment design through two dedicated ground radiation tests. 
The CubeSat was deployed on November 4th, 2024 at 732 km from the Earth using the Polar Satellite Launch Vehicle of the Indian Space Research Organization. 
The polar orbit allows the CubeSat to eventually cover the entire Earth surface over time.
Unlike a sun-synchronous orbit, it also enables different exposure to solar radiation at the 
the same location at different local solar times.
The initial attitude and velocity setting were set to ensure roughly three months of operation. 
The initial estimates were far exceeded as we lost contact with the satellite only at the end of February 2025.

%%%%%LUCA
When the \ac{PDM} accumulates sufficient power, and the \ac{OBC} decides to power the slave $0$ device, the MSP430 MCU  begins the read/write sequence on each of the \ac{NVM} chips.
%When the sequence is complete, we generate a memory dump for each \ac{NVM} chip and hand over the content to the transmission module, which eventually transmits it off to Earth.
The experiments start roughly 10 days after deployment, a time that the mission planner allowed to check mission control systems and data transmission back to Earth. 
Back on the ground, we maintain a reference binary sequence representing the expected memory state after each execution for each \ac{NVM} chip.
%We compare each memory snapshot received from the satellite against this reference to detect errors. 
%The next section illustrates our analysis of the data we received.
We analyze each memory snapshot received from the satellite against this reference to detect errors.

%% file: injection.tex
\section{Error Model}
\label{sec:var}

Our goal is to design a error model capable of accurately representing the experimental data we collect.
Exercising the model allows us to generate a larger set of synthetic error instances useful for developers building space systems using off-the-shelf technology, addressing the practical gap caused by the scarcity of field observations.
%This error model is therefore highly useful for instance in extensive reliability analysis campaigns.

\begin{figure*}[t]
    \centering
    \subfloat[Training and data reconstruction\label{fig:vae_complete}]{
        \includegraphics[width=1.15\columnwidth]{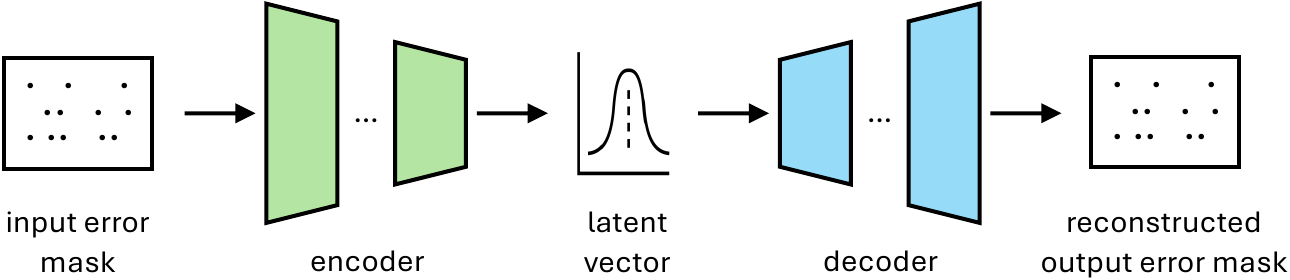}
    } \quad\quad\quad
    \subfloat[Generation of error masks\label{fig:vae_generator}]{
        \includegraphics[width=0.69\columnwidth]{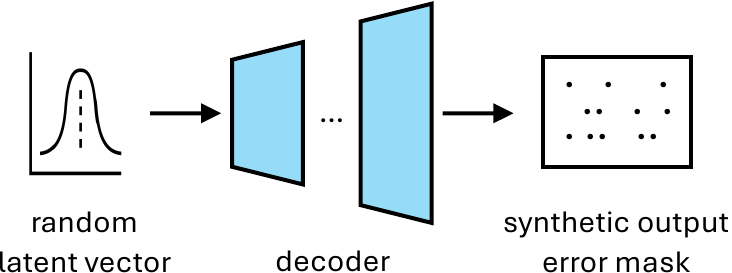}
    } 
          \vspace{-1mm}
  \caption{\ac{VAE} architecture for the error model. \capt{We use the regular encoder-decoder architecture  on the left for training the model with data we collect; then, we feed the sole decoder with random inputs, as shown on the right, and use that as a generator of error masks.}}
    \vspace{-3mm}
    \label{fig:vae}
\end{figure*}

\begin{figure}[tb]
    \centering
        \includegraphics[width=0.36\textwidth]{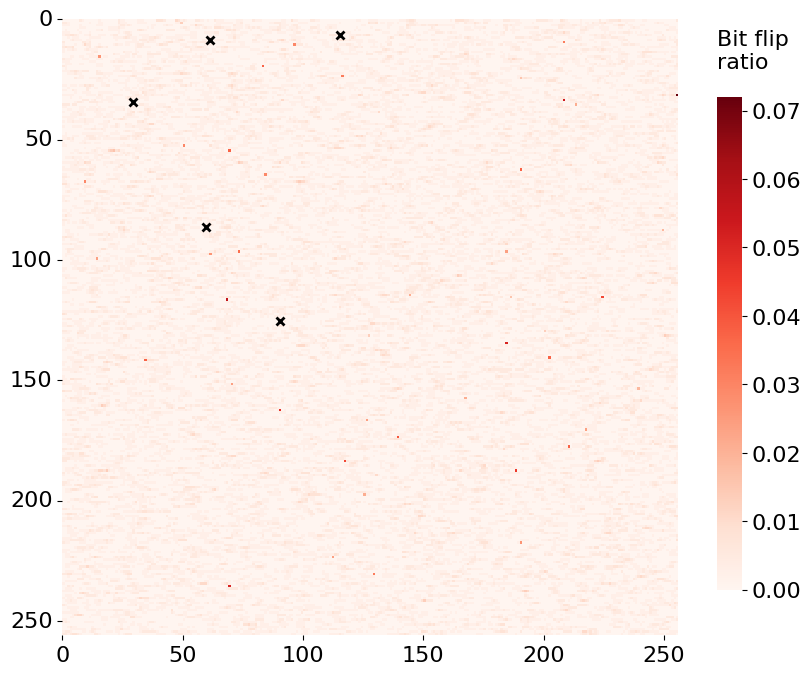}
        \vspace{-1mm}
\caption{Bit error probability heatmap for the \ac{FRAM} chip. \capt{The darker the color of each bit, the higher the observed frequency of bit flip; the \emph{\textbf{X}} mark indicates a bit stuck at a fixed value due to a hard error.}}
    \vspace{-3mm}
    \label{fig:heatmap}
\end{figure}

\subsection{Modeling Challenges}

The operations that produce the bit string are known. Let $S_i\in\{0,1\}^{B}$ denote memory snapshot $i$ and $R_i\in\{0,1\}^{B}$ the corresponding reference bitmap. We compare the two bit strings using
\begin{equation}
  E_i=S_i\oplus R_i, \qquad
  E_{i,b}=1 \iff S_{i,b}\ne R_{i,b}.
  \label{eq:pattern}
\end{equation}
Each one in $E_i$ identifies a bit flip. An \emph{error block} is a sequence of consecutive bit flips; its start bit is the position of the first bit flip and its size is the number of bits flipped starting from that bit. The duration of an error block is the number of consecutive memory snapshots in which it appears. These definitions identify the error patterns in the data we collect; they do not distinguish whether a bit flip is an SEU, an MBU, or a SEFI.

As illustrated in \secref{sec:background}, an error model is usually designed to accurately represent  two aspects: temporal and spatial patterns. 
Modeling temporal patterns is simple. For $N$ memory snapshots, the fraction that contains at least one bit flip is
\begin{equation}
  p=\frac{1}{N}\sum_{i=1}^{N}\mathbf{1}\!\left[\|E_i\|_0>0\right].
  \label{eq:occurrence}
\end{equation}
The value $p$ represents how often the error model generates an error pattern. It does not provide a rate per second, orbit, or radiation dose because the data does not include the exact time or location of every memory snapshot. 

Modeling the  spatial patterns is more challenging, as  we observe multiple errors scattered across the whole memory, as visually exemplified in \figref{fig:heatmap}.
Because of the scarcity of data at hand, visually capturing recurring spatial patterns as in existing literature would be exceedingly difficult.
Unlike existing works, we use deep learning generative techniques to learn spatial patterns from the observed corrupted snapshots and generate new realistic synthetic errors that mimic the original ones.
Candidate techniques in this setting are \acfp{VAE}~\cite{cinelli2021}, \acfp{GAN}~\cite{Creswell2018}, and diffusion models~\cite{Croitoru2023}. 
The latter excel at generating general images, but would be overkill when input data is binary. 
Further, both diffusion models and \acp{GAN}  require large datasets for training~\cite{Creswell2018}.
Applying data augmentation techniques, nevertheless, would not be an option, as even simple operations, like rotations, would alter the semantics of the data, for example, because of the structured organization of memories.
In contrast, \acp{VAE} generally requires limited datasets to operate effectively already, as we also demonstrate next.

A \ac{VAE}~\cite{cinelli2021} is a specific autoencoder neural network that uses a probabilistic model for the internal data representation.
As shown in \figref{fig:vae_complete}, a \ac{VAE} is composed of two symmetric networks, called \emph{encoder} and \emph{decoder}, respectively.
The former takes a high-dimensional input data, the error masks in our case, and transforms it into a compact, low-dimensional representation called \emph{latent vector}.
This encoding maps the input data into a probabilistic distribution, aiming to retain only the most important features of the input while reducing redundancy and noise.
At the right-hand side of \figref{fig:vae_complete}, the decoder starts its processing from the latent vector and reconstructs the original high-dimensional input.

Once the \ac{VAE} is properly trained for encoding and reconstructing the original error patterns, we unplug the encoder; as \figref{fig:vae_generator} shows, the decoder alone, when fed with randomly generated latent vectors, acts as a generator of new synthetic error patterns mimicking the same characteristics of the data used for the training.

\subsection{VAE Error Model}

We prepare input data by extracting \emph{error masks} for each memory snapshot.
An error mask is a binary representation of the memory where a logical value `1' indicates an error bit and a logical `0' represents the absence of an error.
Error masks are simply obtained by XOR-ing each memory snapshot coming off the CubeSat with the reference bitmap. 
Intuitively, an error mask is akin to a bitmap of a common picture.

\fakepar{Design} Due to the limited data available, we design an architecture with a limited number of layers and parameters to avoid overfitting.
The encoder architecture consists of three convolutional layers with increasing filter sizes. 
We select this kind of layers due to their capabilities of learning spatial patterns from input images.
A \ac{ReLU} activation function follows each convolution to speed up convergence and avoid the vanishing gradient in the training process, again motivated by limited data set and the sparsity of inputs.
The output of the last convolution is flattened to feed two parallel dense layers in charge of computing the mean and the log variance of the variables in the latent vector. 
The dense layer connects to a Sigmoid activation function to scale values between 0 and 1, representing the probability of a logical state in the corresponding bit of the error mask.
Finally, we feed outputs into a custom layer implementing the reparametrization trick~\cite{Kingma2015,kingma2022autoencodingvariationalbayes}, which samples the latent vector from a normal distribution parameterized by the learned mean and variance.

The decoder has a dual architecture. 
It takes as input the latent vector and maps it back to the appropriate shape via a dense layer followed by a reshaping pipeline of three convolutional transpose layers. 
The same activation functions as in the encoder layer, that is, the Sigmoid and \ac{ReLU} respectively, follow each layer.
This way, the decoder reconstructs the original input error mask through learned upsampling.
Alternatively, the decoder generates synthetic error masks when the latent vector is sampled randomly.

\fakepar{Training} We train the network by splitting the experimental data we collect from the CubeSat in a 90\% slice for the training set and a 10\% for the validation one. 
We do not split it for a test set due to limited data available.
We select and combine two loss functions that match our specific setting. 
The \ac{WBCE} is essential to handle data sparsity, as error bits represent a limited fraction of a memory snapshot. 
The \ac{KL} divergence term in regularizes the latent vector by forcing the included variables to fit a standard normal distribution. 

The output of the decoder once the \ac{VAE} is trained is not binary, but rather a mask of continuous values between 0 and 1. To map the continuous values to a logical `1' or `0', we use a cutoff threshold $\tau$ that maximizes Youden's J metric~\cite{ruopp2008youden} over the training set.
This is, by construction, the threshold that jointly maximizes sensitivity and specificity. 
We search for such optimal $\tau$ by brute-force down to the third decimal digit, yielding $\tau=0.624$. 
This step is key to enable the decoder to generate realistic and accurate error masks from randomly sampled latent vectors.

\fakepar{Error injection} Let $X$ be application data organized according to the same $B$ bit positions used by the model, and let $G$ be a generated error pattern. We inject the generated bit flips using
\begin{equation}
 X'=X\oplus G.
 \label{eq:inject}
\end{equation}
For every chosen application-level experiment, we first use $p$ to determine whether to inject an error pattern, and then use the model to generate $G$. The XOR operation applies the bit flips to the application data. If an application must reproduce a hard error, the application-level experiment must keep the same error block across consecutive snapshots; the model does not do so on its own.

The consequences depend on the data represented by the affected bits. A flipped least-significant bit in a fixed-point sensor sample or a cached feature may introduce limited noise. A bit flip in a sign or threshold bit may turn a plausible value into an outlier and spuriously trigger or suppress an event. A bit flip in checkpoint control data may steer execution towards an incorrect state. The model supplies error patterns for these experiments; it does not determine the semantics of the data that the application stores.

\section{Validation}
\label{sec:validation}

The preliminary validation we perform is two-pronged.
First, we evaluate the model's capability to reconstruct observed error masks taken from the validation set by performing a k-fold cross-validation ($k=10$); next, we check the model capability to generate synthetic error masks in an accurate way compared to the  dataset we collect from the CubeSat.

In the first case, we take each error mask from the validation set and apply the encoder and decoder stages sequentially.
We compute \emph{precision}, \emph{recall}, and \emph{F1-score} to measure the extent that the encoder captures spatial patterns during training and how those are then reflected in the decoder processing. 
We compute precision and recall from the number of true positives (negatives), that is, when both input and output masks report a logical `1' (`0'), and from the number of false positives (negatives), that is, when the output mask reports a logical `1' (`0') and the original one reports a logical `0' (`1').
We obtain a precision of $0.92$, a recall of $0.91$, resulting in an F1-score of approximately $0.915$.
These values provide evidence that the \ac{VAE} can accurately learn the spatial patterns of bit errors, as seen in the training set.
Note that we were instead unable to spot any such specific pattern visually.

%In the first analysis we take each error mask from the validation set and we apply the encoder and decoder of the designed network in a sequence. 
%We carry out a bitwise comparison of the original error mask and the reconstructed one to count 
%i) the number of bits where both the masks contain 1 -- \ac{TP},
%ii) the number of bits where both the masks contain 0 -- \ac{TN},
%iii) the number of bits where the reconstructed mask contains 1 and the original one 0 -- \ac{FP}, and
%iv) the number of bits where the original mask contains 0 and the reconstructed one 1 -- \ac{FN}.
%We have accumulated the four counters for all items in the validation set and compute the precision ($\frac{TP}{TP+FP}$) and recall ($\frac{TP}{TP+FN}$).
%Results show a precision of \red{XXX} and a recall of \red{XXX}, demonstrating the capability of the model to accurately capture the characteristics of the evaluated dataset. \am{Luca, per utilizzare un'espressione che ho imparato da te: beef up!!!!!!! :D}

Next, we detach the encoder and operate the sole \ac{VAE} decoder by specifying random values for the latent vector used as its input.
To measure how realistic the errors are in the synthetic masks generated this way, we statistically compare them against the entire experimental dataset by computing the Wasserstein distance $W$~\cite{Wasserstein}.
This is a similarity metric between two probability distributions, that, in the discrete domain case, can be seen as the cost of an optimal transport plan to convert one distribution into the other; thus, short distances imply close similarity of the two distributions, while large distances indicate considerable discrepancies.

To compute the Wasserstein distance, we summarize the original dataset and the synthetic one by computing the empirical spatial distribution of the corrupted bits.
We normalize the values of these two distributions compared with the highest value of both distributions.
In this setting, a value $\hat{W} = 0$ indicates that the distributions are identical, a value of  $0 < \hat{W} \ll 1$ indicates that the distributions are very similar and only a small fraction of the maximum possible probability shift is required to move one distribution into the other, whereas a value of $\hat{W} \approx 1$ would indicate that the distributions are maximally different, with their probability mass located at opposite extremes of the support.
Practically, a value of $\hat{W} = 0.02$ indicates that the two distributions differ by only $2\%$ of the maximum possible displacement of probability mass~\cite{Wasserstein}.  

We generate 10,000 synthetic error masks and compute $\hat{W}=0.073$ based on this dataset and compared with the experimental data set. 
This means that the two distributions differ by roughly $7\%$ of the maximum possible displacement of their probability masses. 
Based on this key result, we maintain that, given the scarcity of the data at hand and the inability to obtain more in the short term, the error model we design is already an asset for developers building space systems operating with off-the-shelf hardware.

%% file: discussion.tex
\section{Conclusion}
\label{sec:end}

We developed an error model for a low-power \ac{FRAM} chip deployed on board a CubeSat in \ac{LEO}.
The error model reproduces the effects of physical faults experienced in orbit due to space radiations. 
We use a \ac{VAE} to automatically learn the spatial patterns of bit errors and to synthetically generate additional error patterns mimicking the observed ones. 
The model is capable of producing memory corruption pattern instances for general application-level error simulation experimentation. 
When comparing the model output with the data we collect, our preliminary validation indicates that 0.92 of the generated bit flips correspond to bit flips in the memory snapshots, and the model captures 0.91 of the observed bit flips. 
A comparison of the spatial distribution of bit flips in 10,000 generated error patterns indicates a Wasserstein distance of 0.073; thus,  the two distributions differ roughly $7\%$ of the maximum possible displacement of their probability masses. 

Future work will focus on a more systematic assessment of the error model's accuracy in reproducing the actual error effects observed in orbit. This investigation will  include a visual inspection and comparison of the spatial error patterns exhibited by the synthetically generated and experimentally collected snapshots. Furthermore, we plan to extend the experimental evaluation to different datasets, such as corrupted images produced by image filtering applications, to assess the general applicability  of the approach.

%% file: paper.bib
@inproceedings{zhang2026dynafilter,
  title={DynaFilter: Cloud-driven Dynamic Filtering for Satellite Edge Intelligence},
  author={Zhang, Ziyang and Liu, Jie and Mottola, Luca},
year = {2026},
booktitle = {Proc. Int. Conf. on Mobile Computing and Networking (MobiCom)}
}

@inproceedings{kingma2014vae,
  author = {Kingma, Diederik P. and Welling, Max},
  title = {Auto-Encoding Variational Bayes},
  booktitle = {Proceedings of the International Conference on Learning Representations (ICLR)},
  year = {2014},
  url = {https://arxiv.org/abs/1312.6114}
}

@article{harvesting-survey,
    author = {N. A. Bhatti and M. H. Alizai and A. A. Syed and L. Mottola},
    title = {{Energy Harvesting and Wireless Transfer in Sensor Network Applications: Concepts and Experiences}},
    year = {2016},
    journal = {ACM Trans. on Sensor Networks},
    volume = {12}, 
    number = {3}, 
}

@article{bosser2018single,
  title={Single-event effects in the peripheral circuitry of a commercial ferroelectric random access memory},
  author={Bosser, A. L.  and Gupta, V. and Javanainen, A. and Tsiligiannis, G. and LaLumondiere, S. D. and Brewe, D. and Ferlet-Cavrois, V. and Puchner, H. and Kettunen, H. and Gil, T. and others},
  journal={IEEE Trans. on Nuclear Science},
  volume={65},
  number={8},
  pages={1708--1714},
  year={2018}
}

@article{jesd57,
  year=2023,
  title  = {{Test Procedures for the Measurement of Single-Event Effects in Semiconductor Devices from Heavy-Ion Irradiation}},
  author       = {{JEDEC Solid State Technology Association}},
  number       = {JESD57},
  url          = {https://standards.globalspec.com/std/122027/jedec-jesd-57}
}

@article{ruopp2008youden,
  title={Youden Index and optimal cut-point estimated from observations affected by a lower limit of detection},
  author={Ruopp, M. D.and Perkins, N. J. and Whitcomb, B. W. and Schisterman, E. F.},
  journal={Biometrical Journal: Journal of Mathematical Methods in Biosciences},
  volume={50},
  number={3},
  pages={419--430},
  year={2008}
}

@inproceedings{blackwell2012nanosatellites,
  title={{Nanosatellites for earth environmental monitoring: The MicroMAS project}},
  author={Blackwell, W. and Allen, G. and Galbraith, C. and Hancock, T. and Leslie, R. and Osaretin, I. and Retherford, L. and Scarito, M. and Semisch, C. and Shields, M. and others},
  booktitle={Proc. Specialist Meeting on Microwave Radiometry and Remote Sensing of the Environment (MicroRad)},
  pages={1--4},
  year={2012}
}

@article{esper2000nasa,
  title={{NASA-GSFC nano-satellite technology for earth science missions}},
  author={Esper, J. and Panetta, P. V. and Ryschkewitsch, M. and Wiscombe, W. and Neeck, S.},
  journal={Acta Astronautica},
  volume={46},
  number={2-6},
  pages={287--296},
  year={2000}
}

@article{trusculescu2012nanosatellites,
  title={{Nanosatellites: The Tool for Earth Observation and Near Earth Environment Monitoring}},
  author={Trusculescu, M. and Balan, M. and Dragasanu, C. and Pandele, A. and Piso, M.-I.},
  journal={Earth Observation},
  pages={25--37},
  year={2012}
}

@article{nagel2020nanosatellites,
  title={{Nanosatellites applied to optical Earth observation: a review}},
  author={Nagel, G. Willy and Novo, E. M. de M. and Kampel, M.},
  journal={Revista Ambiente \& {\'A}gua},
  volume={15},
  pages={e2513},
  year={2020}
}

@inproceedings{praks2011aalto,
  title={{Aalto-1-An experimental nanosatellite for hyperspectral remote sensing}},
  author={Praks, J. and Kestil{\"a}, A. and Hallikainen, M. and Saari, H. and Antila, J. and Janhunen, P. and Vainio, R.},
  booktitle={Proc. Int.  Geoscience and Remote Sensing Symp.},
  pages={4367--4370},
  year={2011}
}

@article{schwank2008radiation,
  title={{Radiation effects in MOS oxides}},
  author={Schwank, J. R. and Shaneyfelt, M. R. and Fleetwood, D. M. and Felix, J. A. and Dodd, P. E. and Paillet, P. and Ferlet-Cavrois, V.},
  journal={IEEE Trans. on Nuclear Science},
  volume={55},
  number={4},
  pages={1833--1853},
  year={2008}
}

@article{oldham2003total,
  title={{Total ionizing dose effects in MOS oxides and devices}},
  author={Oldham, T. R. and McLean, F. B.},
  journal={IEEE Trans. on Nuclear Science},
  volume={50},
  number={3},
  pages={483--499},
  year={2003}
}

@inproceedings{yaacoub2025fault,
  title={{Fault Tolerance in Space with Heterogeneous Hardware: Experiences from a 68-day CubeSat Deployment in LEO}},
  author={Yaacoub, A. E. and Voigt, T. and Ruemmer, P. and Mottola, L.},
  booktitle={Proc. Int. Conf. on Embedded Wireless Systems and Networks (EWSN)},
  year={2025},
  pages={1--12}
}

@inproceedings{berck2025launch,
  title={{Launch Fast Learn Fast: Lessons Learned from the R5-S2 and R5-S4 Missions}},
  author={Berck, J. and Pedrotty, S. and Wisbiski, J. and Alexander, M. and Knesek, K. and May, R. and Nunez-Kearny, J. and Stoddard, R. and Kirven, T. and Onsay, R. and others},
  booktitle={Proc. Small Satellite Conf.},
  year={2025},
  pages={1--20}
}

@article{denby2019orbital,
  title={Orbital edge computing: Machine inference in space},
  author={Denby, B. and Lucia, B.},
  journal={IEEE Computer Architecture Letters},
  volume={18},
  number={1},
  pages={59--62},
  year={2019}
}

@inproceedings{orbitalEdge,
  title={Orbital edge computing: Nanosatellite constellations as a new class of computer system},
  author={Denby, B. and Lucia, B.},
  booktitle={Proc. Int. Conf. on Architectural Support for Programming Languages and Operating Systems (ASPLOS)},
  pages={939--954},
  year={2020}
}

@inproceedings{6353715,
  author={Quinn, H. and Graham, P. and Morgan, K. and Baker, Z. and Caffrey, M. and Smith, D. and Bell, R.},
  booktitle={Proc. IEEE Radiation Effects Data Workshop (REDW)}, 
  title={{On-Orbit Results for the Xilinx Virtex-4 FPGA}}, 
  year={2012},
  pages={1-8},

}

@inproceedings{narayana2020hummingbird,
author = {Narayana, S. and Prasad, R. V. and Rao, V. and Mottola, L. and Prabhakar, T. V.},
title = {{Hummingbird: energy efficient GPS receiver for small satellites}},
year = {2020},
booktitle = {Proc. Int. Conf. on Mobile Computing and Networking (MobiCom)}
}

@inproceedings{denby2022tartan,
  title={{Tartan Artibeus: A batteryless, computational satellite research platform}},
  author={Denby, B. and Ruppel, E. and Singh, V. and Che, S. and Taylor, C. and Zaidi, F. and Kumar, S. and Manchester, Z and Lucia, B.},
  booktitle={Proc. Small Satellite Conf.},
  year={2022},
  pages={1--15}
}

@misc{endurosat, 
  url={https://www.endurosat.com/}, 
  title={EnduroSat 1U CubeSat Platform},
  author={EnduroSat},
  year=2025
}

@misc{satnogs, 
  url={https://www.satnogs.org}, 
  title={Open Source Global Network of Satellite Ground-stations},
  year=2025,
  author={SatNOGS}
}

@misc{cesi, 
  url={https://satsearch.co/products/endurosat-1u-cubesat-solar-panel}, 
  title={2 CESI Solar Cells CTJ30},
  year=2025,
  author={satsearch}
}

@article{stakem2004migration,
  title={{Migration of an Image Classification Algorithm to an Onboard Computer for Downlink Data Reduction}},
  author={Stakem, P.},
  journal={Journal of Aerospace Computing, Information, and Communication},
  volume = {1},
  number = {2},
  pages = {108--111},
  year={2004}
}

@inproceedings{1035377,
  author={Lovellette, M. N. and Wood, K. S. and Wood, D. L. and Beall, J. H. and Shirvani, P. P. and Oh, N. and McCluskey, E. J.},
  booktitle={Proc., IEEE Aerospace Conference}, 
  title={{Strategies for fault-tolerant, space-based computing: Lessons learned from the ARGOS testbed}}, 
  year={2002},
  volume={5},
  pages={5--5}
}

@book{pradhan1996, 
editor = {Pradhan, Dhiraj K.}, 
title = {Fault-tolerant computer system design}, 
year = {1996}, 
isbn = {0130578878}, 
publisher = {Prentice-Hall, Inc.}, 
address = {USA} 
}

@ARTICLE{classes,
  author={Bolchini, C. and Cassano, L. and Miele, A. and Toschi, A.},
  journal={IEEE Trans. on Computers}, 
  title={{Fast and Accurate Error Simulation for CNNs Against Soft Errors}}, 
  year={2023},
  volume={72},
  number={4},
  pages={984-997}
}

@ARTICLE{ruospo2022,
  author={Luza, L. M. and Ruospo, A. and Söderström, D. and Cazzaniga, C. and Kastriotou, M. and Sanchez, E. and Bosio, A. and Dilillo, L.},
  journal={IEEE Trans on Emerging Topics in Computing}, 
  title={{Emulating the Effects of Radiation-Induced Soft-Errors for the Reliability Assessment of Neural Networks}}, 
  year={2022},
  volume={10},
  number={4},
  pages={1867-1882}
}

@misc{alen, 
  url={https://alen.space/space-business}, 
  title={How to Do Business ibn Space?},
  author={{Alen}},
  year =2025
}

@article{paikowsky2017,
author = {Paikowsky, D.},
title = {{What Is New Space? The Changing Ecosystem of Global Space Activity}},
journal = {New Space},
volume = {5},
number = {2},
pages = {84--88},
year = {2017}
}

@ARTICLE{binder1975,
  author={Binder, D. and Smith, E. C. and Holman, A. B.},
  journal={IEEE Trans. on Nuclear Science}, 
  title={{Satellite Anomalies from Galactic Cosmic Rays}}, 
  year={1975},
  volume={22},
  number={6},
  pages={2675-2680}
}

@article{Kovar2020,
title = {{Measurement of cosmic radiation in LEO by 1U CubeSat}},
journal = {Radiation Measurements},
volume = {139},
pages = {106471},
year = {2020},
author = {P. Kovar and M. Sommer and D. Matthiae and G. Reitz},
}

@misc{noaa, 
  url={https://www.ncei.noaa.gov/products/space-weather/legacy-data/solar-flares-events}, 
  title={Solar Geophysical Data},
  author={{National Oceanic and Atmospheric Administration}},
  year=2020
}

@misc{spacex, 
  url={https://www.spacex.com/updates/}, 
  title={Geomagnetic storm and recently deployed starlink satellites.},
  author={{SpaceX}},
  year=2025
}

@article{POGHOSYAN201759,
title = {{CubeSat evolution: Analyzing CubeSat capabilities for conducting science missions}},
journal = {Progress in Aerospace Sciences},
volume = {88},
pages = {59--83},
year = {2017},
author = {A. Poghosyan and A. Golkar}
}

@article{Wasserstein,
   author = "Panaretos,  V. M. and Zemel, Y.",
   title = {{Statistical Aspects of Wasserstein Distances}}, 
   journal= "Annual Review of Statistics and Its Application",
   year = "2019",
   volume = "6",
   pages = "405-431"
  }

@misc{kingma2022autoencodingvariationalbayes,
      title={Auto-Encoding Variational Bayes}, 
      author={D. P. Kingma and M. Welling},
      year={2022},
      eprint={1312.6114},
      archivePrefix={arXiv},
      primaryClass={stat.ML},
      url={https://arxiv.org/abs/1312.6114}, 
}

@inproceedings{Kingma2015,
author = {Kingma, D. P. and Salimans, T. and Welling, M.},
title = {Variational dropout and the local reparameterization trick},
year = {2015},
booktitle = {Proc. Int. Conf. on Neural Information Processing Systems (NeurIPS) - Volume 2},
pages = {2575–-2583},
}

@book{cinelli2021,
author = {Cinelli, L. P.},
edition = {1st ed. 2021.},
isbn = {3-030-70679-6},
publisher = {Springer},
title = {Variational methods for machine learning with applications to deep networks },
year = {2021},
}

@ARTICLE{Creswell2018,
  author={Creswell, A. and White, T. and Dumoulin, V. and Arulkumaran, K. and Sengupta, B. and Bharath, A. A.},
  journal={IEEE Signal Processing Magazine}, 
  title={Generative Adversarial Networks: An Overview}, 
  year={2018},
  volume={35},
  number={1},
  pages={53-65}
}

@ARTICLE{Croitoru2023,
  author={Croitoru, F.-A. and Hondru, V. and Ionescu, R. T.  and Shah, M.},
  journal={IEEE Trans. on Pattern Analysis and Machine Intelligence}, 
  title={{Diffusion Models in Vision: A Survey}}, 
  year={2023},
  volume={45},
  number={9},
  pages={10850-10869}
}

@article{korkian2022single,
  author  = {Golnaz Korkian and Daniel Le{\'o}n and
             Francisco J. Franco and Juan C. Fabero and
             Manon Letiche and Yolanda Morilla and
             Pedro Mart{\'i}n-Holgado and Helmut Puchner and
             Hortensia Mecha and Juan A. Clemente},
  title   = {Single Event Upsets Under Proton, Thermal, and
             Fast Neutron Irradiation in Emerging Nonvolatile Memories},
  journal = {IEEE Access},
  volume  = {10},
  pages   = {114566--114585},
  year    = {2022},
  doi     = {10.1109/ACCESS.2022.3217527}
}

@techreport{oldham2013radiation,
  author      = {Timothy R. Oldham and Dakai Chen and
                 Stephen P. Buchner and Kenneth A. LaBel},
  title       = {Radiation Effects Test Guideline Document for
                 Nonvolatile Memories: Lessons Learned},
  institution = {NASA Electronic Parts and Packaging (NEPP) Program},
  year        = {2013},
  month       = apr,
  url         = {https://nepp.nasa.gov/files/24671/Oldham_2013_NVM_Guideline.pdf}
}

@article{dijks2026systematic,
  author  = {Jasper Dijks and Sybren de Jong and
             Alessandra Menicucci and Iklim Akay},
  title   = {Systematic Review of Engineering and Testing
             Approaches for Radiation Hardness Assurance
             in Commercial Space Avionics},
  journal = {Acta Astronautica},
  volume  = {238},
  pages   = {1354--1366},
  year    = {2026},
  doi     = {10.1016/j.actaastro.2025.07.055},
  publisher = {Elsevier}
}

@inproceedings{li2024exploring,
  title={Exploring real-time satellite computing: From energy and thermal perspectives},
  author={Li, Q. and Wang, S. and Xu, C. and Ma, X. and Xu, M. and Zhou, A. and Xing, R. and Yang, B. and Zhu, Z. and Zhang, Y. and others},
  booktitle={2024 IEEE Real-Time Systems Symposium (RTSS)},
  year={2024},
  organization={IEEE}
}

@inproceedings{xing2024deciphering,
  title={Deciphering the enigma of satellite computing with cots devices: Measurement and analysis},
  author={Xing, R. and Xu, M. and Zhou, A. and Li, Q. and Zhang, Y. and Qian, F. and Wang, S.},
  booktitle={Proceedings of the 30th Annual International Conference on Mobile Computing and Networking},
  year={2024}
}

@article{pocketcube,
	author = {Franzese, V.},
	journal = {CEAS Space Journal},
	title ={{PocketQube picosatellites: survey of missions and technologies}},
	year = {2025}
}

@article{ahmed2024internet,
author = {S. Ahmed and B. Islam and K. S. Yildirim and M. Zimmerling and P. Pawe\l{}czak and M. H. Alizai and B. Lucia and L. Mottola and J. Sorber and J. Hester}, 
title = {{The Internet of Batteryless Things}}, 
year = {2024}, 
volume = {67}, 
number = {3}, 
journal = {Commun. ACM}, 
pages = {64–73}
}
